# Transformation From Legal-marking Set to Admissible-marking Set of Petri Nets With Uncontrollable Transitions

ShouGuang Wang, *Senior Member*, *IEEE*, Dan You, MengChu Zhou, *Fellow*, *IEEE,* and Carla Seatsu

*Abstract*—Linear constraint transformation is an essential step to solve the forbidden state problem in Petri nets that contain uncontrollable transitions. This work studies the equivalent transformation from a legal-marking set to its admissible-marking set given such a net. First, the concepts of an escaping-marking set and a transforming marking set are defined. Based on them, two algorithms are given to compute the admissible-marking set and the transforming marking set, which establish the theoretical foundation for the equivalent transformation of linear constraints. Second, the theory about the equivalent transformation of a disjunction of linear constraints imposed to Petri nets with uncontrollable transitions is established. Third, two rules are given to decide the priority of transitions for transformation. Finally, the transformation procedure from a given linear constraint to a logic expression of linear constraints that can describe its entire admissible-marking set is illustrated via two examples.

## Glossary of Symbols

| | |
|---|---|
| $N$ | $\{0, 1, 2, \dots\}$ |
| $N^+$ | $\{1, 2, \dots\}$ |
| $N_k$ | $\{1, 2, \dots, k\}$ |
| $T_c$ | the set of controllable transitions |
| $T_u$ | the set of uncontrollable transitions |
| $N$ | $(P, T, F)$=$(P, T_u \cup T_c, F)$, an ordinary Petri net |
| $\mathcal{M}$ | the set of all possible markings of $N$ |
| $\mathcal{M}_f$ | a forbidden-marking set |
| $\mathcal{L}$ | a legal-marking set |
| $\mathcal{A}$ | the admissible marking set with respect to $\mathcal{L}$ |
| $\mathcal{M}_{WF}$ | the weakly forbidden-marking set with respect to $\mathcal{L}$ |
| $u$ | a supervisory policy |
| $u_{zero}$ | the least permissive supervisory policy that disables all controllable transitions |
| $R(N, m_0)$ | the set of all reachable markings of $N$ from $m_0$ |
| $R_t(N, m)$ | the set of all reachable markings of $N$ from $m$ by firing $t$ only |
| $R(N, m_0, u)$ | the set of all reachable markings under the supervision of a policy $u$ in $N$ from $m_0$ |
| $R(N, m, u_{zero})$ | the set of all reachable markings of $N$ from $m$ by firing uncontrollable transitions only |
| $Q$ | a marking set |
| $\Gamma(Q, t)$ | the escaping-marking set of $Q$ via $t$ |
| $Q_t$ | the transforming marking set of $Q$ via $t$ |
| $m\|_{P'}$ | the restriction of marking $m$ to $P'$ |
| $Q\|_{P'}$ | $\{m\|_{P'} \mid m \in Q\}$, the restriction of the marking set $Q$ to $P'$ |
| $Q^*$ | the all-place-restricted marking set of $Q$ |
| $(\omega, k)$ | a linear constraint |
| $W$ | $\{(\omega_1, k_1), (\omega_2, k_2), \dots, (\omega_n, k_n)\}, n \in N^+$ |
| $\vee(W)$ | the disjunction of the constraints in $W$ |

## I. INTRODUCTION

NOWADAYS, it is critical for industry to seek high resource utilization. Yet it may lead to forbidden states if resources become insufficient. The forbidden states are a kind of states that can reduce the production efficiency, make a great economic loss, and even result in a catastrophic consequence in a safety-critical system. Therefore, how to supervise a discrete event system (DES) to reach no forbidden states during its operation is an important problem in DES control theory, which is called a forbidden state problem.

A Petri net [11], [22], [29], due to its graphical representation and powerful algebraic formulation, has been a popular modeling tool for DES to handle the forbidden state problem. Moreover, most forbidden-state specification that requires a DES to run within a specified set of allowed states can be formalized as a logic expression of linear constraints on the state space of a Petri net model [16], [17], such as the guarantee of deadlock-freedom and liveness of a DES. Note that the markings of Petri net that violate the given linear constraints are forbidden ones and those satisfy them are legal ones. Note that

This work is in part supported by National Natural Science Foundation of China under Grant 61472361, 61100056 and 61374148, Zhejiang Provincial Natural Science Foundation for Distinguished Young Scholars (No. LR14F020001), Zhejiang Sci. & Tech. Project under Grant 2013C31111, Zhejiang NNST Key Laboratory under Grant 2013E10012 , and Zhejiang Gongshang University Innovation Project under Grant 3080XJ2513233 and 3080XJ2513236.

S. G. Wang and D. You are with School of Information & Electronic Engineering, Zhejiang Gongshang University, Hangzhou, 310018, China (corresponding author: 0571-28877734; e-mail: wsg5000@hotmail.com, youdan000@hotmail.com ).
M. C. Zhou is with the Department of Electrical and Computer Engineering, New Jersey Institute of Technology, Newark, NJ 07102, USA (e-mail: zhou@njit.edu).
C. Seatzu is with the Department of Electrical and Electronic Engineering, University of Cagliari, 09123 Cagliari, Italy (e-mail: seatzu@diee.unica.it).

legal markings here are not related to initial marking of a Petri net but only given constraints. Thus they may not be the reachable markings as required in some deadlock control research [8]. A forbidden state problem can also be described as how to ensure the behavior of the plant net to conform to the given linear constraints.

In a Petri net model, transitions correspond to events in a DES. If an event cannot be prevented from occurring, its corresponding transition is called an uncontrollable one. Otherwise, it is called a controllable one. Petri nets can thereby be divided into ones with and without uncontrollable transitions.

For a Petri net without uncontrollable transitions, the key to solving the forbidden state problem is to synthesize a supervisor that can enforce the given linear constraints on the reachable markings of the plant net. The supervisor synthesis techniques were investigated in many early studies [10], [12], [14], [28].

As for a Petri net with uncontrollable transitions, the problem is rather complex since the uncontrollable transitions cannot be prevented from firing by any external supervisor and thus a forbidden marking may be reached from a legal one with the firing of uncontrollable transitions only. Actually, in order to make such a Petri net satisfy the forbidden-state specification, its behavior should be restricted within the admissible-marking set instead of the legal-marking set [12]. Note that an admissible marking is a marking from which no forbidden ones can be reached by firing uncontrollable transitions only [15]. This implies that the given linear constraints have to be transformed into a logic expression of linear constraints that can describe the admissible-marking set and then a supervisor can be designed to enforce the transformation result on the plant net. Therefore, the study on linear constraint transformation becomes an essential task to solve the forbidden state problem for Petri nets with uncontrollable transitions.

Generally speaking, methods applicable for constraint transformation can be divided into ones with and without the analysis on reachability set. The former ones mainly proceed as follows: First, the admissible-marking set of a given legal-marking set is identified from the reachability graph. Next, suitable and compact constraints are found to express the admissible-marking set. Methods in [2]-[8], [10] all belong to the former type. It is clear that these methods can hardly be applied to large-sized nets since they suffer from the state explosion problem. Hence, many efforts have been made on the latter ones as summarized next.

Moody and Antsaklis [24] propose the concept of an admissible linear constraint. The markings satisfying it are all admissible ones. They present a method that can transform a given linear constraint that is inadmissible into an admissible one and then a supervisor is designed to enforce the admissible constraint on the plant net. The supervisor is computationally efficient and simple in structure but usually not maximally permissive (optimal). This is because the obtained admissible constraint just describes a subset of the admissible-marking set, or the proposed transformation is not an equivalent one. Note that the equivalent transformation [16], [20] requires that no admissible marking is removed.

Basile *et al.* [1] improve the method in [24] by adding two parameters to the matrix containing the uncontrollable columns of the plant incidence matrix. As a result, a larger subnet of the admissible-marking set can be derived and thus the designed supervisor has higher behavioral permissiveness. Besides, Iordache and Antsaklis [13] also improve the method in [24] by using the concepts of firing vector and Parikh vector. Unfortunately, none of their results can describe the entire admissible-marking set, i.e., their transformations are non-equivalent ones.

Some work is done for some subclasses of Petri nets with uncontrollable transitions. Uzam [25] proves the existence of an optimal monitor-based supervisor for a class of Petri nets with uncontrollable transitions and two different optimal supervisors are designed for a Petri net example. For Petri nets whose uncontrollable influence subnets are forward synchronization and forward conflict free (FSFCF) nets [18], forward synchronization and backward conflict free (FSBCF) nets [19] and forward concurrent free (FCF) nets [20], Luo *et al.* present equivalent constraint transformation and techniques for synthesizing the optimal supervisor based on a crux path set. For Petri nets whose uncontrollable influence subnets are FSBCF nets, Wang *et al.* [26] propose a new method, which has higher computational efficiency than that in [19], to equivalently transform a given linear constraint into an admissible one and then an optimal monitor-based supervisor can be designed for the admissible constraint.

As for Petri nets with general structures, [17], [21], [23] point out it is impossible to equivalently transform some linear constraints into admissible ones. Luo *et al.* [17] thereby propose the concept of a weakly admissible constraint. Based on it, they claim that they propose an algorithm that can equivalently transform a linear constraint into a disjunction of weakly admissible ones and an optimal supervisor is designed for the weakly admissible ones. However, a counterexample given later in this paper reveals that it is a non-equivalent transformation only. Therefore, how to directly transform a given linear constraint equivalently into a logic expression of linear constraints that can describe the entire admissible marking set of the given linear constraint without the analysis on reachability set remains open. This paper aims to solve this long-standing difficult problem. The new contributions of this paper include:

1) Escaping-marking sets and transforming marking sets are defined and their related properties are given;

2) A theoretical framework is established to compute the admissible-marking set given a legal-marking set and the transforming marking set via an uncontrollable transition given a union of two marking sets;

3) A counterexample is presented to reveal that the constraint transformation in [17] is not an equivalent one and the reason for its non-equivalence is stated;

4) Equivalent transformation of a disjunction of linear constraints via an uncontrollable transition without the analysis on reachability set is presented, which corrects the fault in [17]; and

5) Two rules are proposed for equivalently transforming a given linear constraint into a logic expression of linear constraints that can describe its entire admissible-marking set.

The remainder of this paper is organized as follows. Section II gives the related notions of Petri nets. Section III establishes a theoretical framework of the transformation from a legal-marking set to its admissible-marking set. Section IV shows the equivalent transformation of a disjunction of linear constraints and two rules to decide the priority of transitions for transformation. Section V provides two examples to illustrate the equivalent transformation. Finally, Section VI concludes this paper.

## II. PRELIMINARIES

An ordinary Petri net is a 3-tuple $N=(P, T, F)$ where $P$ and $T$ are finite, nonempty, and disjoint sets. $P$ is a set of places, and $T$ is a set of transitions. $F \subseteq (P \times T) \cup (T \times P)$ is a set representing all the flow relations. Given a net $N=(P, T, F)$ and a node $x \in P \cup T$, $^\bullet x = \{y \in P \cup T | (y, x) \in F\}$ is the preset of $x$, while $x^\bullet = \{y \in P \cup T | (x, y) \in F\}$ is the postset of $x$. $\forall X \subseteq P \cup T$, $^\bullet X = \bigcup_{x \in X} {}^\bullet x$ and $X^\bullet = \bigcup_{x \in X} x^\bullet$. The incidence matrix of $N$ is denoted by $[N]$: $P \times T \to \{-1, 0, 1\}$ indexed by $P$ and $T$ such that $[N](p, t)=-1$ if $p \in {}^\bullet t \backslash t^\bullet$; $[N](p, t)=1$ if $p \in t^\bullet \backslash {}^\bullet t$; otherwise $[N](p, t)=0$, $\forall p \in P$ and $\forall t \in T$.

A *marking* or *state* of a Petri net $N=(P, T, F)$ is a mapping $m : P \to N$ where $N=\{0, 1, 2, \ldots\}$. In general, we use the multi-set notation $\sum_{p \in P} m(p)p$ to denote vector $m$, where $m(p)$ denotes the number of tokens in place $p$ at $m$. For example, $m=[2, 1, 0, 0]^T$ is denoted by $m=2p_1+p_2$. $p$ is marked by $m$ if $m(p)>0$. The set of all possible markings of $N$ is defined as $\mathcal{M}= N^{|P|}$. $(N, m_0)$ is called a net system or marked net given its initial marking $m_0$.

A transition $t$ is enabled at a marking $m$, denoted by $m [t>$, if $\forall p \in {}^\bullet t$, $m(p)>0$. An enabled transition $t$ at $m$ can fire, resulting in $m'$, denoted by $m [t>m'$, where $m'(p)=m(p)+[N](p, t)$. A sequence of transitions $\alpha=t_1 t_2 \ldots t_k$, $t_i \in T$, $i \in N_k=\{1, 2, \ldots, k\}$, is fireable from $m$ if $m_i [t_i > m_{i+1}$, $i \in N_k$, where $m_1=m$. In such a case, we use $m [\alpha > m_{k+1}$ to denote that $m_{k+1}$ is reachable from $m$ after firing $\alpha$. Let $R(N, m_0)$ denote the set of all reachable markings of $N$ from $m_0$.

A transition without any input place is called a source transition, and one without any output place is called a sink transition. Note that a source transition is unconditionally enabled and its firing generates tokens but consumes no token. Firing a sink transition consumes tokens but does not produce any. A place without any input transition is called a source place, and one without any output transition is called a sink place.

The transition set $T$ is partitioned into two disjoints subsets: $T_u$ is the set of uncontrollable ones, and $T_c$ is the set of controllable ones. A controllable transition can be prevented from firing by a supervisory policy, but uncontrollable transitions cannot.

The set of reachable markings under the supervision of a policy $u$ in $N$ from $m_0$ is denoted by $R(N, m_0, u)$. The least permissive supervisory policy, denoted as $u_{zero}$, disables all controllable transitions. $R(N, m_0, u_{zero})$ is the set of markings uncontrollably reachable from $m_0$, where all controllable ones are disabled. Clearly, it is the smallest one, i.e., $R(N, m_0, u_{zero}) \subseteq R(N, m_0, u)$ for any $u$.

## III. TRANSFORMATION FROM A LEGAL-MARKING SET TO ITS ADMISSIBLE-MARKING SET

A forbidden-state specification for a DES requires that the system never enters a specified set of forbidden states, or, equivalently, that the system remains in a specified set of allowed states [12]. The sets of forbidden states and allowed states are called the forbidden-marking set and the legal-marking set, denoted by $\mathcal{M}_F$ and $\mathcal{L}$, respectively. Clearly, $\mathcal{M}_F \subseteq \mathcal{M}$, $\mathcal{L} \subseteq \mathcal{M}$, and $\mathcal{L}=\mathcal{M}\text{-}\mathcal{M}_F$.

There is no need to supervise such a Petri net if $R(N, m_0) \subseteq \mathcal{L}$ since all the reachable markings satisfy the control specification. Hence, we do not study such a case in this paper. In the remaining discussion, we assume that $(N, m_0)$ is a net system containing uncontrollable transitions such that $R(N, m_0) \not\subset \mathcal{L}$.

For a Petri net with uncontrollable transitions, since a forbidden marking may be reached from a legal one by firing uncontrollable transitions, its behavior has to be restricted within a smaller marking set instead of the legal-marking set, which is called the admissible-marking set defined as follows.

*Definition* 1: Let $Q \subseteq \mathcal{M}$ be a marking set. $A(Q)=\{m \in Q | R(N, m, u_{zero}) \subseteq Q\}$ is called the *admissible-marking set* with respect to $Q$ and $M_{WF}(Q)= \{m \in Q | \exists m' \in R(N, m, u_{zero}), m' \in \bar{Q}\}$ is called the *weakly forbidden-marking set* with respect to $Q$, where $\bar{Q}$ is a complement of $Q$, i.e., $\mathcal{M}\text{-}Q$.

In this paper, given a legal-marking set $\mathcal{L}$ for a net, we use $\mathcal{A}$ to denote $A(\mathcal{L})$ and $\mathcal{M}_{WF}$ to denote $M_{WF}(\mathcal{L})$. The relationship among all the defined marking sets is given in Fig. 1.

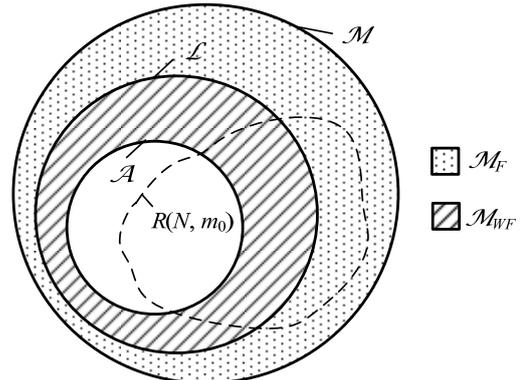

Fig. 1. Relationship among the defined marking sets

According to Definition 1, a legal-marking set $\mathcal{L}$ can be separated into two subsets, namely the admissible-marking set $\mathcal{A}$ and the weakly forbidden-marking set $\mathcal{M}_{WF}$. The markings in $\mathcal{A}$ can never reach a forbidden marking by firing uncontrollable transitions only; while those in $\mathcal{M}_{WF}$ can. Therefore, although the markings in $\mathcal{M}_{WF}$ are legal ones, they are the markings that a Petri net should be prevented from reaching.

*Definition* 2 [12]: Let $\mathcal{L} \subseteq \mathcal{M}$ be a legal-marking set, $\mathcal{A}$ be the admissible marking set with respect to $\mathcal{L}$, and $u$ be a control policy. The policy $u$ is called optimal (maximally permissive) if $R(N, m_0, u) = R(N, m_0) \cap \mathcal{A}$.

*Definition* 3: Let $Q_1$ and $Q_2 \subseteq \mathcal{M}$ be two marking sets. $Q_1$ and $Q_2$ are called equivalent if $A(Q_1) = A(Q_2)$, denoted as $Q_1 \cong Q_2$.

### A. Escaping-marking Set and Transforming Marking Set

We use $R_t(N, m)$ to denote the set of all reachable markings (including $m$) of $N$ from $m$ by firing $t$ only. For example, given $m = (0, 2, 3)^T$ for the net $N$ in Fig. 2, we have $R_t(N, m) = \{(0, 2, 3)^T, (1, 1, 2)^T, (2, 0, 1)^T\}$.

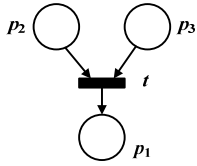

Fig. 2. A simple Petri net

*Definition* 4: Let $Q \subseteq \mathcal{M}$ be a marking set for $N$ and $t$ be an uncontrollable transition. $\Gamma(Q, t) = \{m \in Q \mid \exists\, m' \in R_t(N, m), m' \in \bar{Q}\}$ is called the *escaping-marking set* of $Q$ via $t$. $Q$-$\Gamma(Q, t)$ is denoted by $Q_t$, called the *transforming marking set* of $Q$ via $t$.

*Property* 1: $Q_t \subseteq Q$ and $Q_t = \{m \in Q \mid R_t(N, m) \subseteq Q\}$.

$\Gamma(Q, t)$ denotes a subset of $Q$, where the markings can reach a marking outside of $Q$ by firing uncontrollable transition $t$ only, i.e., the markings can "run away from" $Q$ after firing $t$. Accordingly, $Q_t$ denotes a marking set in which no marking can "run away from" $Q$, or to be exact, no marking can "run away from" $Q_t$ if only $t$ is allowed to fire.

Consider the Petri net in Fig. 2 with $t$ being uncontrollable. Consider $Q = \{(1, 0, 0)^T, (1, 0, 1)^T, (0, 1, 1)^T, (1, 1, 1)^T, (0, 2, 2)^T\}$. By Definition 4, we have $\Gamma(Q, t) = \{(1, 1, 1)^T, (0, 2, 2)^T\}$ since $(2, 0, 0)^T$ is reachable from them by firing $t$ and $(2, 0, 0)^T \notin Q$. Accordingly, $Q_t = \{(1, 0, 0)^T, (1, 0, 1)^T, (0, 1, 1)^T\}$.

*Property* 2: $\forall\, t \in T_u$, $\Gamma(\mathcal{A}, t) = \varnothing$ and $\mathcal{A} = \mathcal{A}$.

Note that we use $Q_{tt'}$ to denote $Q_t$-$\Gamma(Q_t, t')$ where $t$ and $t' \in T_u$, i.e., $Q_{tt'} = (Q_t)_{t'}$. Similarly, $Q_{\alpha\beta\ldots\gamma} = (\ldots((Q_\alpha)_\beta)\ldots)_\gamma$ where $\alpha$, $\beta$, …, $\gamma \in T_u$. Here, $\alpha$, $\beta$, …, $\gamma$ may include a same uncontrollable transition multiple times.

*Property* 3: $\Gamma(Q_t, t) = \varnothing$ and $Q_{tt} = Q_t$.

The following properties are straightforward from Definitions 1 and 4.

*Property* 4: $\Gamma(Q, t) \subseteq \mathcal{M}_{WF}$ if $\mathcal{A} \subseteq Q \subseteq \mathcal{L}$.

*Property* 5: $\mathcal{A} \subseteq Q_t \subseteq \mathcal{L}$ if $\mathcal{A} \subseteq Q \subseteq \mathcal{L}$.

Given a Petri net $(N, m_0)$ with a legal-marking set $\mathcal{L}$, $\Gamma(\mathcal{L}, t)$ can denote the set of legal markings which can reach forbidden ones by firing $t$ only according to Definition 4. As a result, $\mathcal{L}$-$\Gamma(\mathcal{L}, t)$ is denoted by $\mathcal{L}_t$, in which no marking can reach a forbidden one by firing $t$ only. Moreover, we have $\mathcal{L}_{\alpha\beta\ldots\gamma} = (\ldots((\mathcal{L}_\alpha)_\beta)\ldots)_\gamma$ as stated above.

*Theorem* 1: $\mathcal{L} \supseteq \mathcal{L}_\alpha \supseteq \mathcal{L}_{\alpha\beta} \supseteq \ldots \supseteq \mathcal{L}_{\alpha\beta\ldots\gamma} \supseteq \mathcal{A}$

*Proof*: It is trivial from Properties 1 and 5. ∎

The relationship among the marking sets in Theorem 1 is shown as Fig. 3.

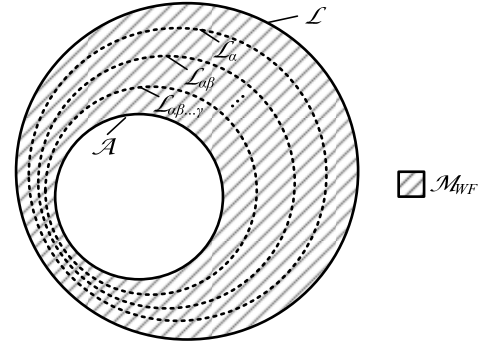

Fig. 3. Relationship among $\mathcal{L}$, $\mathcal{L}_\alpha$, $\mathcal{L}_{\alpha\beta}$, …, $\mathcal{L}_{\alpha\beta\ldots\gamma}$ and $\mathcal{A}$

*Theorem* 2: Let $Q = \mathcal{L}_{\alpha\beta\ldots\gamma}$. $Q = \mathcal{A}$ if $\forall\, t \in T_u$, $\Gamma(Q, t) = \varnothing$.

*Proof*: According to Theorem 1, $\mathcal{A} \subseteq Q \subseteq \mathcal{L}$ holds. By contradiction, suppose that $Q \neq \mathcal{A}$. Clearly, we have $Q \cap \mathcal{M}_{WF} \neq \varnothing$, i.e., $\exists\, m \in Q$, $\exists\, m' \in R(N, m, u_{zero})$, $m' \in \mathcal{M}_F$. Hence, $\exists\, t \in T_u$, $\Gamma(Q, t) \neq \varnothing$, which contradicts the condition that $\forall\, t \in T_u$, $\Gamma(Q, t) = \varnothing$. Therefore, $Q = \mathcal{A}$ holds. ∎

*Theorem* 3: Given a marking set $Q \subseteq \mathcal{M}$ and $t \in T_u$, $Q \cong Q_t$.

*Proof*: It is obvious that $A(Q) = A(Q_t)$. Hence, $Q \cong Q_t$ holds. ∎

*Theorem* 4: $\mathcal{L} \cong \mathcal{L}_\alpha \cong \mathcal{L}_{\alpha\beta} \cong \ldots \cong \mathcal{L}_{\alpha\beta\ldots\gamma}$.

*Proof*: Straightforward from Theorem 3. ∎

### B. Computation of Admissible and Transforming Marking Sets

Given a marking $m$ for $N$ and $P' \subseteq P$, we use $m|_{P'}$ to denote the restriction of marking $m$ to $P'$. For example, $m = (1, 2, 3)^T$ is a marking for the net in Fig. 2, we have $m|_{\{p1\}} = 1$ and $m|_{\{p1, p2\}} = (1, 2)^T$. Given a marking set $Q \subseteq \mathcal{M}$ and $P' \subseteq P$, $Q|_{P'} = \{m|_{P'} \mid m \in Q\}$ is called the restriction of the marking set $Q$ to $P'$. For example, $Q = \{m \in \mathcal{M} \mid 2m(p_1) + m(p_2) \leq 2\}$ is a marking set for the net in Fig. 2. We have $Q|_{\{p1\}} = \{0, 1\}$, $Q|_{\{p3\}} = N$, and $Q|_{\{p1, p2\}} = \{(0, 0)^T, (0, 1)^T, (0, 2)^T, (1, 0)^T\}$.

*Definition* 5: Given a marking set $Q \subseteq \mathcal{M}$ and $p \in P$, $p$ is called an *unrestricted place* if $Q|_{\{p\}} = N$; otherwise, a *restricted place*. The set of all restricted places with respect to $Q$ is denoted by $P_Q$. The *all-place-restricted marking set* of $Q$ is defined as $Q^* = Q|_{P_Q}$.

For example, $Q = \{m \in \mathcal{M} \mid 2m(p_1) + m(p_2) \leq 2\}$ is a marking set for the net in Fig. 2. We have $P_Q = \{p_1, p_2\}$ and $Q^* = Q|_{P_Q} = \{(0, 0)^T, (0, 1)^T, (0, 2)^T, (1, 0)^T\}$.

In the following discussion, we assume that for any legal-marking set $\mathcal{L}$, its all-place-restricted marking set $\mathcal{L}^*$ is a finite marking set, i.e., it contains finite markings only.

**Algorithm 1:** Computation of an admissible-marking set

**Input:** An ordinary Petri net $(N, m_0)$ and a legal-marking set $\mathcal{L}$.

**Output:** A marking set $Q$.

1)   $Q=\mathcal{L}$;
2)   **while** $\exists t \in T_u$, $\Gamma(Q, t) \neq \varnothing$ **do**
3)       $Q=Q_t$;
4)   **end while**
5)   **Output:** $Q$;
6)   **End.**

Algorithm 1 presents the procedure to obtain the admissible-marking set of a legal-marking set, which is performed by iteratively computing the transformation marking set via each uncontrollable transition.

*Theorem* 5: Given an ordinary Petri net $(N, m_0)$ and a legal-marking set $\mathcal{L}$ as the inputs, Algorithm 1 can output a marking set $Q=\mathcal{A}$.

*Proof:* In order to prove that Algorithm 1 can output a marking set $Q$, we have to prove that Steps 2 to 4 of Algorithm 1 are executed finite times. Since $\mathcal{L}^*$ is a finite marking set and the number of places in a Petri net is limited, a marking set $Q' \subseteq \mathcal{L}$ can be obtained after executing Steps 2 to 4 finite times, which satisfies that $Q'^*$ is a finite marking set and the number of restricted places cannot be increased any more in spite of the execution of Steps 2 to 4. Next, every time Steps 2 to 4 are executed, several markings are removed from the finite marking set $Q'^*$. Clearly, Steps 2 to 4 can be executed for finite times only. Since Algorithm 1 can output $Q$, we have $Q=\mathcal{A}$ due to Theorem 2. ∎

For a marking set $Q=Q_1 \cup Q_2$, Algorithm 2 is presented to compute $Q_t$ under the condition that $(Q_1)_t$ and $(Q_2)_t$ are known, which also reveals the relationship between $Q_t$ and $(Q_1)_t \cup (Q_2)_t$. Note that $Q_1^*$ and $Q_2^*$ are required to be finite marking sets in Algorithm 2. To make the following section easy to understand, we provide a case of $Q=Q_1 \cup Q_2$ in Fig. 4, where $Q_1 \cap Q_2 \neq \varnothing$.

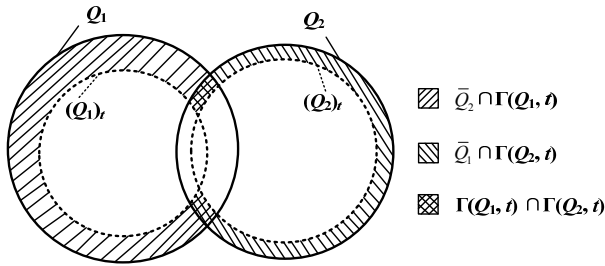

Fig. 4. A case of $Q=Q_1 \cup Q_2$

**Algorithm 2:** Computation of $Q_t$
**Input:** A marking set $Q=Q_1 \cup Q_2$ for $(N, m_0)$ and $t \in T_u$.
**Output:** A marking set $Q_{out}$.

1)   $B_1=(Q_1)_t$;
2)   $B_2=(Q_2)_t$;
3)   $C_1=\{m \in (Q_1\text{-}B_1) \cap \bar{B}_2 | \exists m' \in R_t(N, m) \cap Q_1$ such that $m'[t \rhd m''$ and $m'' \in \bar{Q}_1 \cap B_2\}$;
4)   $C_2=\{m \in (Q_2\text{-}B_2) \cap \bar{B}_1 | \exists m' \in R_t(N, m) \cap Q_2$ such that $m'[t \rhd m''$ and $m'' \in \bar{Q}_2 \cap B_1\}$;
5)   **while** $C_1 \cup C_2 \neq \varnothing$ **do**
6)       $B_1=B_1 \cup C_1$;
7)       $B_2=B_2 \cup C_2$;
8)       $C_1=\{m \in (Q_1\text{-}B_1) \cap \bar{B}_2 | \exists m' \in R_t(N, m) \cap Q_1$ such that $m'[t \rhd m''$ and $m'' \in \bar{Q}_1 \cap B_2\}$;
9)       $C_2=\{m \in (Q_2\text{-}B_2) \cap \bar{B}_1 | \exists m' \in R_t(N, m) \cap Q_2$ such that $m'[t \rhd m''$ and $m'' \in \bar{Q}_2 \cap B_1\}$;
10)   **end while**
11)   $Q_{out}=B_1 \cup B_2$;
12)   **Output:** $Q_{out}$;
13)   **End.**

Algorithm 2 presents how to compute the transforming marking set of a union of two marking sets via an uncontrollable transition under the condition that the transforming marking set of each given marking set is known.

*Theorem* 6: Given a marking set $Q=Q_1 \cup Q_2$ for $(N, m_0)$ and $t \in T_u$, Algorithm 2's output $Q_{out}=Q_t$.

*Proof:* The proof includes two parts, namely, $Q_t \supseteq Q_{out}$, and $Q_t \subseteq Q_{out}$.

First, we prove that $Q_t \supseteq Q_{out}$. According to Property 1, it is obvious that $B_1 \subseteq Q_t$ and $B_2 \subseteq Q_t$ in Steps 1 and 2, i.e., the markings in both $B_1$ and $B_2$ can never "run away from" $Q_1 \cup Q_2$ by firing $t$. $C_1$ denotes a subset of $(Q_1\text{-}B_1) \cap \bar{B}_2$, where once the markings "run away from" $Q_1$ by firing $t$, they enter $B_2$. Since the markings in $B_2$ can never "run away from" $Q_1 \cup Q_2$ by firing $t$, $C_1 \subseteq Q_t$. Similarly, $C_2 \subseteq Q_t$. Then, since the markings in $C_1$ are added to $B_1$ and those in $C_2$ are added to $B_2$, as stated in Steps 6 and 7, we still have $B_1 \subseteq Q_t$ and $B_2 \subseteq Q_t$. With the repeated execution of Steps 5 to 10, $B_1$ and $B_2$ are expanded and $B_1 \subseteq Q_t$ and $B_2 \subseteq Q_t$ always hold. When $B_1$ and $B_2$ can never be expanded, $Q_{out}=B_1 \cup B_2$. Hence, $Q_t \supseteq Q_{out}$ holds.

Next, we prove that $Q_t \subseteq Q_{out}$. By contradiction, suppose that $Q_t \not\subset Q_{out}$. Then there exists a marking $m \in Q_t \cap \bar{Q}_{out}$. Since $m \in Q_t$, we have $R_t(N, m) \subseteq Q_1 \cup Q_2$. Moreover, since $m \in \bar{Q}_{out}$, we accordingly have the following three cases:

1) $m \in \bar{Q}_{out} \cap Q_1 \cap \bar{Q}_2$;
2) $m \in \bar{Q}_{out} \cap Q_2 \cap \bar{Q}_1$; and
3) $m \in \bar{Q}_{out} \cap Q_1 \cap Q_2$.

Case 1: It is easy to know that $\bar{Q}_{out} \cap Q_1 \subseteq \Gamma(Q_1, t)$. Thus, $m \in \bar{Q}_2 \cap \Gamma(Q_1, t)$. Clearly, $m$ can reach a marking outside of $Q_1$ by firing $t$ only. Since $R_t(N, m) \subseteq Q_1 \cup Q_2$, we can conclude that once the net evolves from $m$ to $m' \in \bar{Q}_1$, $m' \in Q_2$ holds. Moreover, we have $m' \in \bar{Q}_1 \cap \Gamma(Q_2, t)$ since otherwise $m \in Q_{out}$. For the same reason, the net can evolve from $m'$ to a marking outside of $Q_2$ by firing $t$ only, and once a marking $m'' \in \bar{Q}_2$ is reached from $m'$, we have $m''\in \bar{Q}_2\cap\Gamma(Q_1, t)$. As a result, we can conclude that $t$ can continuously fire infinite times from $m$.

Case 2: Similar to Case 1. We can conclude that $t$ can continuously fire infinite times from $m$.

Case 3: It is easy to know that $m\in\Gamma(Q_1, t)\cap\Gamma(Q_2, t)$. Since $m\in\Gamma(Q_2, t)$, we know that $m$ can reach a marking outside of $Q_2$ by firing $t$ only. Since $R_t(N, m)\subseteq Q_1\cup Q_2$, we can conclude that once a marking $m'\in\bar{Q}_2$ is reached from $m$, $m'\in Q_1$ holds. Moreover, we have $m'\in\bar{Q}_2\cap\Gamma(Q_1, t)$ since otherwise $m\in Q_{out}$. As discussed in Case 1, since $m'\in\bar{Q}_2\cap\Gamma(Q_1, t)$, we can conclude that $t$ can continuously fire infinite times from $m$.

Cases 1 to 3 imply that $\Gamma(Q_1, t)\neq\varnothing$ and $\Gamma(Q_2, t)\neq\varnothing$, and since $t$ can continuously fire infinite times, we can know that $\exists p\in t^\bullet$, $p$ is a restricted place both under $Q_1$ and $Q_2$ and the number of tokens in $p$ can increase infinitely with the firing of $t$. This means both $Q_1{}^*$ and $Q_2{}^*$ are infinite sets, which contradicts the fact that they are finite. Hence, $Q_t\subseteq Q_{out}$.

Therefore, $Q_t=Q_{out}$ holds since $Q_t\supseteq Q_{out}$ and $Q_t\subseteq Q_{out}$. ∎

*Remark:* Algorithm 2 and Theorem 6 indicate that for a marking set $Q=Q_1\cup Q_2$, $Q_t\supseteq (Q_1)_t\cup(Q_2)_t$ holds. Moreover, some markings in $\Gamma(Q_1, t)\cup\Gamma(Q_2, t)$ also belong to $Q_t$ since although these markings can "run away from" $Q_1$ or $Q_2$ by firing $t$ only, they can never "run away from" $Q_1\cup Q_2$ by firing $t$ only. More specifically, if only $t$ can fire, once they "run away from" $Q_1$, they enter $Q_2$, and once they "run away from" $Q_2$, they enter $Q_1$. Note that these markings are what Algorithm 2 aims to compute in an iteration way.

To illustrate Algorithm 2, let $Q=Q_1\cup Q_2$ be a marking set for the net in Fig. 2, where $Q_1=\{m_1\text{-}m_{10}\}=\{(1, 0, 0)^T, (0, 1, 1)^T, (4, 0, 0)^T, (3, 1, 1)^T, (1, 1, 2)^T, (1, 1, 1)^T, (0, 2, 2)^T, (1, 2, 2)^T, (2, 1, 1)^T, (0, 1, 2)^T\}$ and $Q_2=\{m_4, m_8, m_{11}\text{-}m_{16}\}=\{(3, 1, 1)^T, (1, 2, 2)^T, (2, 0, 0)^T, (2, 0, 1)^T, (3, 0, 0)^T, (0, 2, 3)^T, (0, 3, 3)^T, (0, 3, 2)^T\}$.

First, it is clear that $B_1=(Q_1)_t=\{m_1\text{-}m_4\}$ and $B_2=(Q_2)_t=\{m_{11}\text{-}m_{13}\}$. Since the markings $m_5\text{-}m_9$ in $(Q_1\text{-}B_1)\cap\bar{B}_2$ satisfy that once they "run away from" $Q_1$ by firing $t$ only, they "enter" $(Q_2)_t$, we have $C_1=\{m_5\text{-}m_9\}$. For the similar reason, we have $C_2=\varnothing$. Hence, $B_1$ can be expanded with the result being $B_1=\{m_1\text{-}m_9\}$ and $B_2$ is still equal to $\{m_{11}\text{-}m_{13}\}$. Next, we have $C_1=\varnothing$ and $C_2=\{m_{14}, m_{15}\}$ since $m_{14}, m_{15}\in(Q_2\text{-}B_2)\cap\bar{B}_1$ satisfies that once it "runs away from" $Q_2$ by firing $t$ only, it "enters" $B_1$. Hence, $B_1$ is still equal to $\{m_1\text{-}m_9\}$ and $B_2=\{m_{11}\text{-}m_{15}\}$ after being expanded. Now, we have $C_1=\varnothing$ and $C_2=\varnothing$. Therefore, $B_1$ and $B_2$ cannot be expanded and the final result is $Q_t= B_1\cup B_2 =\{m_1\text{-}m_9, m_{11}\text{-}m_{15}\}$.

The reachability analysis of all the markings in $Q$ is shown in Fig. 5, which intuitively verifies the correctness of the result.

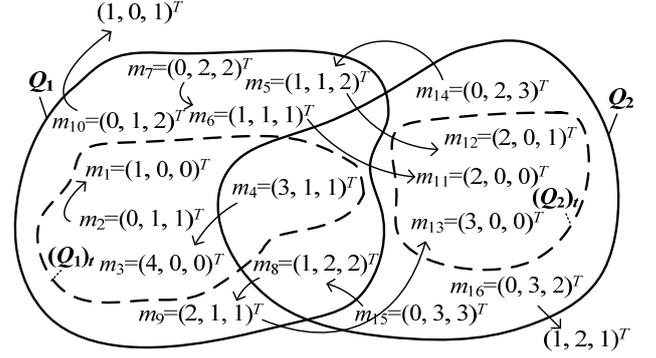

Fig. 5. Reachability analysis of markings in $Q$

*Remark of Section III:* A forbidden-state specification is usually formalized as constraints. It is known that constraint forms are various. Hence, this section deals with the problem of how to obtain the admissible-marking set of a legal-marking set by the analysis on reachability set. Actually, it establishes a theoretical framework, which reveals the nature of equivalent constraint transformation and thus can provide ideas for the study on direct transformation approach of constraints in specific form. Under the theoretical framework of this section, Section IV studies the equivalent transformation of a special form of constraints, i.e., a disjunction of linear constraints.

## IV. LINEAR CONSTRAINT TRANSFORMATION

In this section, the forbidden-state specification for a DES is formalized as a logic expression of linear constraints for its Petri net model. In other words, we use a logic expression of linear constraints to describe the legal-marking set.

A linear constraint $(\omega, k)$ requires the markings $m$ of a Petri net to satisfy $\omega\bullet m\leq k$ where $k$ is an integer and $\omega$ is a weight vector from $P$ to $N$. Let $W=\{(\omega_1, k_1), (\omega_2, k_2), \dots, (\omega_n, k_n)\}$, $n\in N^+$ denote a set of linear constraints. The disjunction of the constraints in $W$ is denoted as $\vee(W)$, that is, $\vee_{(\omega, k)\in W}\omega\bullet m\leq k$.

We use $Q_{(\omega, k)}=\{m\in\mathcal{M}\mid\omega\bullet m\leq k\}$ to denote a marking whose elements meet a linear constraint $(\omega, k)$, and $Q_{\vee(W)}=\cup_{(\omega, k)\in W}Q_{(\omega, k)}$ to denote one whose elements meet the disjunction of linear constraints in $W$. Clearly, for any $(\omega, k)$ in this paper, $Q^*_{(\omega, k)}$ is a finite marking set.

*Definition 6* [20]: Given a Petri net $(N, m_0)$ with a linear constraint $(\omega, k)$, the weight of transitions is defined as a row vector, that is, $\varpi=\omega\bullet[N]$.

The following properties are straightforward from Definition 6.

*Property 6:* Let $Q_{(\omega, k)}$ be a marking set for $(N, m_0)$ and $t\in T_u$ such that $\varpi(t)\geq 0$. $\forall m\in\bar{Q}_{(\omega, k)}$, $R_t(N, m)\subseteq\bar{Q}_{(\omega, k)}$.

*Property 7:* Let $Q_{(\omega, k)}$ be a marking set for $(N, m_0)$ and $t\in T_u$ such that $\varpi(t)\leq 0$. $\forall m\in Q_{(\omega, k)}$, $R_t(N, m)\subseteq Q_{(\omega, k)}$.

Note that two logic expressions of linear constraints are called equivalent if the marking sets described by them are equivalent.

### A. Existing Constraint Transformation Method [17]

Luo *et. al.* [17] give an algorithm to equivalently transform a given linear constraint into a disjunction of weakly admissible ones. However, the obtained disjunction of weakly admissible constraints describes a marking set that may be just a subset of the admissible-marking set of the given linear constraint. In other words, the transformation method in [17] is not an equivalent one [27], contrary to what they claim in [17].

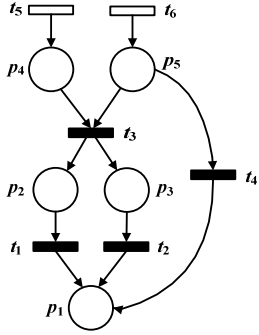

Fig. 6. A counterexample

For example, $(\omega, 3)$: $m(p_1)+m(p_2)+m(p_3) \leq 3$ is a linear constraint for the net in Fig. 6 with $t_1$- $t_4$ being uncontrollable. The constraint transformation procedure according to Algorithm 1 [17] is as follows:

1) $(\omega, 3)$ is transformed via $t_3$ into

$(\omega_1, 3)$: $m(p_1)+m(p_2)+m(p_3)+2m(p_4) \leq 3$ ∨
$(\omega_2, 3)$: $m(p_1)+m(p_2)+m(p_3)+2m(p_5) \leq 3$

2) $(\omega_1, 3)$ is transformed via $t_4$ into

$(\omega_3, 3)$: $m(p_1)+m(p_2)+m(p_3)+2m(p_4)+m(p_5) \leq 3$

Since $(\omega_2, 3)$ and $(\omega_3, 3)$ are both weakly admissible linear constraints, the transformation result is $(\omega_2, 3) \vee (\omega_3, 3)$.

Consider a marking $m=\{0, 0, 0, 1, 2\}^T$. It is easy to know that $m$ is an admissible marking for $(\omega, 3)$ since $\forall m' \in R(N, m, u_{zero})$, $m'(p_1)+m'(p_2)+m'(p_3) \leq 3$. However, $m$ does not satisfy $(\omega_2, 3) \vee (\omega_3, 3)$. Hence, this example shows that the constraint transformation [17] is not an equivalent one.

The reason that the transformation [17] is non-equivalent is that Algorithm 1 [17] is presented based on the equation:

$$A(Q_{\vee(W)}) = \cup_{(\omega, k) \in W} A(Q_{(\omega, k)}). \qquad (1)$$

On the premise that (1) holds, a disjunction of two or more linear constraints can be equivalently transformed by equivalently transforming each constraint independently via the transformation method in [17]. However, (1) may not be correct in some cases. Consider $(\omega_1, 3) \vee (\omega_2, 3)$ in the above counterexample. We have $\{0, 0, 0, 1, 2\}^T \in A(Q_{(\omega_1, 3) \vee (\omega_2, 3)})$ but $\{0, 0, 0, 1, 2\}^T \notin A(Q_{(\omega_1, 3)}) \cup A(Q_{(\omega_2, 3)})$ by the reachability analysis of firing uncontrollable transitions only.

Actually, we can conclude that $A(Q_{\vee(W)}) \supseteq \cup_{(\omega, k) \in W} A(Q_{(\omega, k)})$ holds. This is because that there may exist some markings in $Q_{\vee(W)}$ satisfying that they can "run away from" a marking set $Q_{(\omega, k)}$ by firing uncontrollable transitions but can never "run away from" $Q_{\vee(W)}$. Clearly, these markings are admissible ones for $Q_{\vee(W)}$. However, they are unfortunately lost during the transformation procedure in [17].

Therefore, for a disjunction of two or more linear constraints, all the constraints should be considered as a whole for the equivalent transformation instead of transforming them independently.

In what follows, we first present the equivalent transformation of a linear constraint via an uncontrollable transition under the theoretical framework of Section III, and then propose a method for equivalently transforming a disjunction of two or more linear constraints via an uncontrollable transition. In addition, two rules are given to decide the priority of transitions for transformation.

### B. Equivalent Transformation of Linear Constraint via Uncontrollable Transition

*Definition* 7: The uncontrollable transition gain transformation (UTGT) function is $\rho$: $\Omega \times T_u \times P \rightarrow \Omega$, where $\Omega$ is the set of all linear-constraints. It is defined as $\forall (\omega, k) \in \Omega$, $\forall t \in T_u$, $\forall p \in P$, $(\omega', k') = \rho((\omega, k), t, p)$ and we have

$$
\begin{cases}
k' = k \\
\forall p' \in P, \quad \omega'(p') = \begin{cases} \omega(p') & p' \neq p \vee p' \notin {}^\bullet t \\ \omega(p') + \varpi(t) & p' = p \wedge p' \in {}^\bullet t \setminus t^\bullet \\ k+1 & p' = p \wedge p' \in {}^\bullet t \cap t^\bullet \end{cases}
\end{cases}.
$$

*Definition* 8: Given an uncontrollable transition $t$ and a linear constraint $(\omega, k)$, $\varrho((\omega, k), t)$ is defined as

$$
\varrho((\omega, k), t) = \begin{cases} \{(\omega, k)\} & \varpi(t) \leq 0 \\ \cup_{p \in {}^\bullet t} \{\rho((\omega, k), t, p)\} & \varpi(t) > 0 \end{cases},
$$

where $\rho$ is defined in Definition 7.

The concepts of $\rho((\omega, k), t, p)$ and $\varrho((\omega, k), t)$ in this work are different from those defined in [17]. Note that only in a case that $p$ is both the input and output place of $t$, the result of $\rho((\omega, k), t, p)$ in this work is different from that in [17]. The concept of $\varrho((\omega, k), t)$ now considers the case $\varpi(t) \leq 0$ but not in [17].

For example, $(\omega, 1)$: $m(p_1) \leq 1$ is a linear constraint for the net in Fig. 7, where $t$ is an uncontrollable transition. Clearly, $\rho((\omega, 1), t, p_2) = (\omega', 1)$: $m(p_1)+2m(p_2) \leq 1$ in this work, while $\rho((\omega, 1), t, p_2) = (\omega'', 1)$: $m(p_1)+m(p_2) \leq 1$ in [17]. They are different.

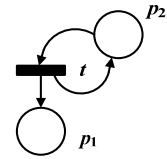

Fig. 7. A Petri net with $(\omega, 1)$: $m(p_1) \leq 1$

*Property* 8: Let $(\omega', k) = \rho((\omega, k), t, p)$ where $p \in {}^\bullet t \setminus t^\bullet$, $\varpi'(t) = 0$ if $\varpi(t) > 0$.

*Property* 9: Let $(\omega', k) = \rho((\omega, k), t, p)$ where $p \in {}^\bullet t \cap t^\bullet$, $\varpi'(t) = \varpi(t)$ if $\varpi(t) > 0$.

*Lemma* 1: Given a linear constraint $(\omega, k)$ and an uncontrollable transition $t$ with $\varpi(t) > 0$, we have $\forall m \in Q_{(\omega', k)}$, $R_t(N, m) \subseteq Q_{(\omega', k)}$, where $(\omega', k) \in \varrho((\omega, k), t)$.

*Proof*: We have the following two cases:

1) $(\omega', k) = \rho((\omega, k), t, p)$, where $p \in {}^\bullet t \setminus t^\bullet$

We have $\varpi'(t)=0$ by Property 8. Hence, $\forall m \in Q_{(\omega', k)}$, $R_t(N, m) \subseteq Q_{(\omega', k)}$.

2) $(\omega', k)=\rho((\omega, k), t, p)$, where $p \in {}^\bullet t \cap t^\bullet$

Since $\omega'(p)=k+1$ by Definition 7, it is easy to see that $\forall m \in Q_{(\omega', k)}$, $m(p)=0$. Since $p \in {}^\bullet t$, $t$ cannot fire. Clearly, $\forall m \in Q_{(\omega', k)}$, $R_t(N, m) \subseteq Q_{(\omega', k)}$.

Therefore, the conclusion holds.  ∎

*Theorem* 7: $(Q_{(\omega, k)})_t = Q_{\vee(W)}$, where $W=\varphi((\omega, k), t)$.

*Proof:* We have two cases: $\varpi(t) \leq 0$ and $\varpi(t) > 0$.

1) $\varpi(t) \leq 0$

Straightforward from Property 7, we have $(Q_{(\omega, k)})_t = Q_{(\omega, k)}$.

2) $\varpi(t) > 0$

First, we prove that $(Q_{(\omega, k)})_t \supseteq Q_{\vee(W)}$. $\forall (\omega', k) \in W$, we have $\omega' \geq \omega$ according to Definitions 7 and 8. It can be inferred that $\forall m$, $\omega' \bullet m \leq k$, we have $\omega \bullet m \leq k$. That is to say, any marking that satisfies some linear constraint in $W$ must satisfies $(\omega, k)$, i.e., $Q_{(\omega, k)} \supseteq Q_{\vee(W)}$ holds. According to Lemma 1, it is clear that $\forall m \in Q_{\vee(W)}$, $R_t(N, m) \subseteq Q_{\vee(W)}$. Hence, we have $(Q_{(\omega, k)})_t \supseteq Q_{\vee(W)}$.

Next, we prove that $(Q_{(\omega, k)})_t \subseteq Q_{\vee(W)}$. By contradiction, suppose that there exists a marking $m \in (Q_{(\omega, k)})_t$ satisfying $m \notin Q_{\vee(W)}$, i.e.,

$$\forall (\omega', k) \in W, \ \omega' \bullet m > k. \tag{2}$$

Let $\alpha$ be a sequence of transitions that consists of only $t$ and can fire from $m$. $|\alpha|$ is finite since otherwise $m \notin (Q_{(\omega, k)})_t$. Moreover, it is clear that there exists $\alpha$ such that $m[\alpha > m_z$ and $\exists p \in {}^\bullet t$, $m_z(p)=0$. Since $m \in (Q_{(\omega, k)})_t$, we have $\omega \bullet m_z \leq k$. According to Definitions 7 and 8, there exists $(\omega_z, k) \in W$ such that $\omega_z \bullet m_z = \omega \bullet m_z$. Since $\omega \bullet m_z \leq k$, we have $\omega_z \bullet m_z \leq k$. Since $\omega_z \bullet m_z = \omega_z \bullet m + \omega_z(t) \bullet |\alpha|$ and $\varpi_z(t) \geq 0$ due to Properties 8 and 9, we have $\omega_z \bullet m_z \geq \omega_z \bullet m$. Since $\omega_z \bullet m_z \leq k$, which contradicts (2). Hence, $(Q_{(\omega, k)})_t \subseteq Q_{\vee(W)}$.

Therefore, $(Q_{(\omega, k)})_t = Q_{\vee(W)}$.  ∎

### C. Equivalent Transformation of Disjunction of Linear Constraints via Uncontrollable Transition

Theorem 7 reveals that a linear constraint can be equivalently transformed via an uncontrollable transition by Definition 8. For a disjunction of linear constraints, the equivalent transformation via an uncontrollable transition is next presented.

*Theorem* 8: Let $Q=Q_{(\omega1, k1)} \cup Q_{(\omega2, k2)}$ be a marking set for $(N, m_0)$ and $t \in T_u$. $(Q)_t=(Q_{(\omega1, k1)})_t \cup (Q_{(\omega2, k2)})_t$ if $\varpi_1(t) \geq 0$ and $\varpi_2(t) \geq 0$.

*Proof:* Let $Q=Q_{(\omega1, k1)} \cup Q_{(\omega2, k2)}$ and $t$ be the inputs of Algorithm 2. First, we have $B_1=(Q_{(\omega1, k1)})_t$ and $B_2=(Q_{(\omega2, k2)})_t$. Next, it is easy to see that $C_1 \subseteq \Gamma(Q_{(\omega1, k1)}, t) \cap \overline{(Q_{(\omega2, k2)})_t}$. Here, we use $D$ to denote $\Gamma(Q_{(\omega1, k1)}, t) \cap \overline{(Q_{(\omega2, k2)})_t}$. Clearly, we have $D=D_1 \cup D_2$, where $D_1=\Gamma(Q_{(\omega1, k1)}, t) \cap \overline{Q}_{(\omega2, k2)}$ and $D_2=\Gamma(Q_{(\omega1, k1)}, t) \cap \Gamma(Q_{(\omega2, k2)}, t)$. 1) Consider $m \in D_1$. By Property 6, we have $\forall m \in D_1$, $R_t(N, m) \subseteq \overline{Q}_{(\omega2, k2)}$ since $\varpi_2(t) \geq 0$. Hence, $\forall m \in D_1$, $R_t(N, m) \subseteq \overline{(Q_{(\omega2, k2)})_t}$; 2) Consider $m \in D_2$. We can see that

$\forall m \in D_2$, once the net evolves from $m$ to $m' \in \overline{D}_2$, $m' \in D_1 \cup D_3$ holds, where $D_3=\Gamma(Q_{(\omega2, k2)}, t) \cap \overline{Q}_{(\omega1, k1)}$. Hence, it is easy to conclude that $\forall m \in D_2$, $R_t(N, m) \subseteq \overline{(Q_{(\omega2, k2)})_t}$. As a result, we have $\forall m \in D$, $R_t(N, m) \subseteq \overline{(Q_{(\omega2, k2)})_t}$. This implies that $C_1=\varnothing$. Similarly, we have $C_2=\varnothing$. Finally, we have $Q_{out}=(Q_{(\omega1, k1)})_t \cup (Q_{(\omega2, k2)})_t$. According to Theorem 6, we have $Q_t=Q_{out}=(Q_{(\omega1, k1)})_t \cup (Q_{(\omega2, k2)})_t$.  ∎

*Corollary* 1: Let $W=\{(\omega_1, k_1), (\omega_2, k_2), …, (\omega_n, k_n)\}$, $n \in N^+$ be a set of linear constraints for $(N, m_0)$ and $t \in T_u$. $(Q_{\vee(W)})_t = \cup_{(\omega, k) \in W} (Q_{(\omega, k)})_t$ if $\varpi_i(t) \geq 0$, $\forall i \in \{1, 2, …, n\}$.

*Theorem* 9: Let $Q=Q_{(\omega1, k1)} \cup Q_{(\omega2, k2)}$ be a marking set for $(N, m_0)$ and $t \in T_u$. $Q_t=Q_{(\omega1, k1)} \cup Q_{(\omega2, k2)}$ if $\varpi_1(t) \leq 0$ and $\varpi_2(t) \leq 0$.

*Proof:* Let $Q=Q_{(\omega1, k1)} \cup Q_{(\omega2, k2)}$ and $t$ be the inputs of Algorithm 2. First, we have $B_1=(Q_{(\omega1, k1)})_t$ and $B_2=(Q_{(\omega2, k2)})_t$, as stated in Steps 1 and 2. Since $\varpi_1(t) \leq 0$ and $\varpi_2(t) \leq 0$, $B_1=Q_{(\omega1, k1)}$ and $B_2=Q_{(\omega2, k2)}$. Clearly, $C_1=\varnothing$ and $C_2=\varnothing$. As a result, the output $Q_{out}=Q_{(\omega1, k1)} \cup Q_{(\omega2, k2)}$. According to Theorem 6, we have $Q_t=Q_{out}=Q_{(\omega1, k1)} \cup Q_{(\omega2, k2)}$.  ∎

*Corollary* 2: Let $W=\{(\omega_1, k_1), (\omega_2, k_2), …, (\omega_n, k_n)\}$, $n \in N^+$ be a set of linear constraints for $(N, m_0)$ and $t \in T_u$. $(Q_{\vee(W)})_t = Q_{\vee(W)}$ if $\varpi_i(t) \leq 0$, $\forall i \in \{1, 2, …, n\}$.

*Definition* 9: Given two linear constraints $(\omega_i, k_i)$ and $(\omega_j, k_j)$ for a Petri net $(N, m_0)$ and $t \in T_u$, $C_{i \rightarrow j}=\{m \in \Gamma(Q_{(\omega i, ki)}, t) \cap \overline{Q}_{(\omega j, kj)}$ $| \exists m' \in R_t(N, m) \cap Q_{(\omega i, ki)}$ such that $m'[t > m''$ and $m'' \in \overline{Q}_{(\omega i, ki)} \cap Q_{(\omega j, kj)}\}$ is called the *complementary-marking set* from $(\omega_i, k_i)$ to $(\omega_j, k_j)$.

*Property* 10: $C_{i \rightarrow j}=\varnothing$ if $\varpi_i(t) \leq 0$ or $\varpi_j(t) \geq 0$.

Under the condition that $\varpi_i(t) > 0$ and $\varpi_j(t) < 0$, the complementary-marking set $C_{i \rightarrow j}$ describes the markings in $\Gamma(Q_{(\omega i, ki)}, t) \cap \overline{Q}_{(\omega j, kj)}$, which satisfy that once they reach a marking outside of $Q_{(\omega i, ki)}$ by firing $t$, the reachable marking belongs to $Q_{(\omega j, kj)}$, or in other words, once they "run away from" $Q_{(\omega i, ki)}$ by firing $t$, they "enter" $Q_{(\omega j, kj)}$.

*Property* 11: Given two linear constraints $(\omega_i, k_i)$ and $(\omega_j, k_j)$ for a Petri net $(N, m_0)$ and $t \in T_u$ with $\varpi_i(t) > 0$ and $\varpi_j(t) < 0$, $C_{i \rightarrow j}$ can be described by the following logic expression of linear constraints:

$$\vee_{\lambda=1}^{n} \Delta_\lambda, \tag{3}$$

where $\Delta_\lambda$ represents

$$\begin{cases} \forall p \in {}^\bullet t, m(p) \geq \lambda \\ \omega_j \square m > k_j \\ \omega_i \square m + (\lambda-1)\varpi_i(t) \leq k_i \\ \omega_i \square m + \lambda \varpi_i(t) > k_i \\ \omega_j \square m + \lambda \varpi_j(t) \leq k_j \end{cases}$$

and $n \in N^+$, $n \leq k_i / \varpi_i(t)+1$.

*Proof:* $\Delta_1$ deals with the markings in $\overline{Q}_{(\omega j, kj)} \cap Q_{(\omega i, ki)}$ satisfying that a) they can reach a marking outside of $Q_{(\omega i, ki)}$ after firing $t$ once and b) the reachable marking belongs to $Q_{(\omega j,}$ $_{kj}$), $\Delta_2$ deals with the markings in $\overline{Q}_{(\omega j,\ kj)} \cap Q_{(\omega i,\ ki)}$ satisfying that a) they can reach a marking outside of $Q_{(\omega i,\ ki)}$ after firing $t$ twice and b) the reachable marking belongs to $Q_{(\omega j,\ kj)}$, and so forth. Hence, it is clear that (3) describes the markings in $\Gamma(Q_{(\omega i,\ ki)},\ t) \cap \overline{Q}_{(\omega j,\ kj)}$, which satisfy that once they reach a marking outside of $Q_{(\omega i,\ ki)}$ by firing $t$, the reachable marking belongs to $Q_{(\omega j,\ kj)}$. ∎

In what follows, $C_{i \rightarrow j}$ can also be used to denote (3) for simplification.

As shown in Fig. 8, there are two linear constraints $(\omega_i,\ k_i)$ and $(\omega_j,\ k_j)$ satisfying that $Q_{(\omega i,\ ki)} \cap Q_{(\omega j,\ kj)} \neq \varnothing$ and an uncontrollable transition $t$ such that $\varpi_i(t) > 0$ and $\varpi_j(t) < 0$. Note that $m_1$ is a marking that satisfies $\Delta_1$ in (3) while $m_1'$ is not such a marking. Moreover, $m_2$ and $m_2'$ satisfy $\Delta_2$, and $m_3$ satisfies $\Delta_3$.

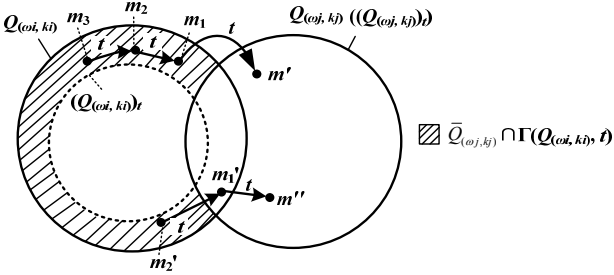

Fig. 8. $\varpi_i(t) > 0$ and $\varpi_j(t) < 0$

*Theorem* 10: Let $Q = Q_{(\omega 1,\ k1)} \cup Q_{(\omega 2,\ k2)}$ be a marking set for $(N, m_0)$ and $t \in T_u$. $Q_t = (Q_{(\omega 1,\ k1)})_t \cup Q_{(\omega 2,\ k2)} \cup C_{1 \rightarrow 2}$ if $\varpi_1(t) > 0$ and $\varpi_2(t) < 0$.

*Proof*: Let $Q = Q_{(\omega 1,\ k1)} \cup Q_{(\omega 2,\ k2)}$ and $t$ be the inputs of Algorithm 2. While $B_1 = (Q_{(\omega 1,\ k1)})_t$ and $B_2 = Q_{(\omega 2,\ k2)}$ after execution of Steps 1 and 2, we have $C_1 = \{m \in \Gamma(Q_{(\omega 1,\ k1)},\ t) \cap \overline{Q}_{(\omega 2,\ k2)} \mid \exists\ m' \in R_t(N, m)\ \cap Q_{(\omega 1,\ k1)}$ such that $m''[t > m''$ and $m'' \in \overline{Q}_{(\omega 1,\ k1)} \cap Q_{(\omega 2,\ k2)}\}$ and $C_2 = \varnothing$. Clearly, $C_1$ is exactly the complementary-marking set $C_{1 \rightarrow 2}$. While $C_1 \neq \varnothing$, Steps 5 to 10 can run again. Then, we have $B_1 = (Q_{(\omega 1,\ k1)})_t \cup C_{1 \rightarrow 2}$ and $B_2$ is still equal to $Q_{(\omega 2,\ k2)}$. Here, we use $C_1'$ and $C_2'$ to denote $C_1$ and $C_2$ in the second execution. Since $B_2$ is not expanded, it can be inferred that $C_1' = \varnothing$. Clearly, $C_2' = \varnothing$. Therefore, we have $Q_{out} = (Q_{(\omega 1,\ k1)})_t \cup Q_{(\omega 2,\ k2)} \cup C_{1 \rightarrow 2}$. It is clear that $Q_t = (Q_{(\omega 1,\ k1)})_t \cup Q_{(\omega 2,\ k2)} \cup C_{1 \rightarrow 2}$ due to Theorem 6. ∎

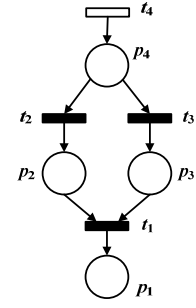

Fig. 9. A Petri net with $Q$: $m(p_1) + m(p_3) \leq 1 \vee m(p_1) + m(p_2) + m(p_4) \leq 1$

Consider the net in Fig. 9 with $t_1$-$t_3$ being uncontrollable. Let $Q = Q_{(\omega 1,\ 1)} \cup Q_{(\omega 2,\ 1)}$ be a marking set for the net, where

$(\omega_1,\ 1)$: $m(p_1) + m(p_3) \leq 1$ and
$(\omega_2,\ 1)$: $m(p_1) + m(p_2) + m(p_4) \leq 1$.

Clearly, $\varpi_1(t_3) = 1 > 0$ and $\varpi_2(t_3) = -1 < 0$. According to Theorem 10, we have $Q_{t3} = (Q_{(\omega 1,\ 1)})_{t3} \cup Q_{(\omega 2,\ 1)} \cup C_{1 \rightarrow 2}$ where

$(Q_{(\omega 1,\ 1)})_{t3} = Q_{(\omega 1',\ 1)},$　$(\omega_1',\ 1)$: $m(p_1) + m(p_3) + m(p_4) \leq 1$ and

$$C_{1 \rightarrow 2}: \begin{cases} m(p_4) \geq 1 \\ m(p_1) + m(p_2) + m(p_4) > 1 \\ m(p_1) + m(p_3) \leq 1 \\ m(p_1) + m(p_3) + 1 > 1 \\ m(p_1) + m(p_2) + m(p_4) - 1 \leq 1 \end{cases}$$

$$\vee \begin{cases} m(p_4) \geq 2 \\ m(p_1) + m(p_2) + m(p_4) > 1 \\ m(p_1) + m(p_3) + 1 \leq 1 \\ m(p_1) + m(p_3) + 2 > 1 \\ m(p_1) + m(p_2) + m(p_4) - 2 \leq 1 \end{cases}.$$

The first conjunction of linear constraints of $C_{1 \rightarrow 2}$ describes the markings in $\Gamma(Q_{(\omega 1,\ 1)},\ t3) \cap \overline{Q}_{(\omega 2,\ 1)}$ that "run away from" $Q_{(\omega 1,\ 1)}$ but "enter" $Q_{(\omega 2,\ 1)}$ after firing $t3$ once, and the second one describes those that "run away from" $Q_{(\omega 1,\ 1)}$ after firing $t3$ twice and "enter" $Q_{(\omega 2,\ 1)}$.

Here, $C_{1 \rightarrow 2}$ can be reduced into:

$$\begin{cases} m(p_4) \geq 1 \\ m(p_1) + m(p_3) = 1 \\ m(p_1) + m(p_2) + m(p_4) = 2 \end{cases}$$

$$\vee \begin{cases} m(p_4) \geq 2 \\ m(p_1) + m(p_3) = 0 \\ 1 < m(p_1) + m(p_2) + m(p_4) \leq 3 \end{cases}.$$

*Theorem* 11: Let $Q = Q_{(\omega 1,\ k1)} \cup Q_{(\omega 2,\ k2)}$ be a marking set for $(N, m_0)$ and $t \in T_u$. $Q_t = (Q_{(\omega 1,\ k1)})_t \cup (Q_{(\omega 2,\ k2)})_t \cup C_{1 \rightarrow 2} \cup C_{2 \rightarrow 1}$.

*Proof*: Straightforward from Theorems 8 to 10. ∎

Note that Theorem 11 is a combination of Theorems 8 to 10. It presents the result of $Q_t$ no matter what the weights of $t$ for two linear constraints are. Clearly, according to Theorem 11, we have

$$Q_t = \begin{cases} (Q_{(\omega1,k1)})_t \cup (Q_{(\omega2,k2)})_t & \varpi_1(t) \geq 0 \wedge \varpi_2(t) \geq 0 \\ Q_{(\omega1,k1)} \cup Q_{(\omega2,k2)} & \varpi_1(t) \leq 0 \wedge \varpi_2(t) \leq 0 \\ (Q_{(\omega1,k1)})_t \cup Q_{(\omega2,k2)} \cup C_{1\rightarrow2} & \varpi_1(t) > 0 \wedge \varpi_2(t) < 0 \\ Q_{(\omega1,k1)} \cup (Q_{(\omega2,k2)})_t \cup C_{2\rightarrow1} & \varpi_1(t) < 0 \wedge \varpi_2(t) > 0 \end{cases}.$$

*Corollary* 3: Let $W = \{(\omega_1, k_1), (\omega_2, k_2), ..., (\omega_n, k_n)\}$, $n \in N^+$ be a set of linear constraints for $(N, m_0)$ and $t \in T_u$. $(Q_{\vee(W)})_t = \cup_{(\omega, k) \in W} (Q_{(\omega, k)})_t \cup_{(i, j) \in E} C_{i\rightarrow j}$, where $E = \{(i, j) | i, j \in \{1, 2, ..., n\}$, $\varpi_i(t) > 0$ and $\varpi_j(t) < 0\}$.

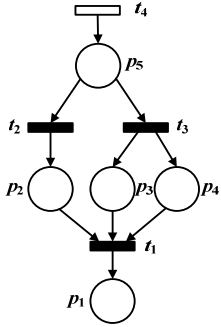

Fig. 10. A Petri net with $Q$: $m(p_1)+m(p_2)+m(p_5) \leq 3 \vee m(p_1)+m(p_3) \leq 3 \vee m(p_1)+m(p_4) \leq 3$

Consider the net in Fig. 10 with $t_1$-$t_3$ being uncontrollable. Let $Q = Q_{(\omega1, 3)} \cup Q_{(\omega2, 3)} \cup Q_{(\omega3, 3)}$ be a marking set for the net, where

$(\omega_1, 3)$: $m(p_1)+m(p_2) + m(p_5) \leq 3$,

$(\omega_2, 3)$: $m(p_1)+m(p_3) \leq 3$ and

$(\omega_3, 3)$: $m(p_1)+m(p_4) \leq 3$.

Since $\varpi_1(t_3) < 0$, $\varpi_2(t_3) > 0$ and $\varpi_3(t_3) > 0$, we have $Q_{t3} = Q_{(\omega1, 3)} \cup Q_{(\omega2, 3)} \cup Q_{(\omega3, 3)})_{t3} \cup C_{2\rightarrow1} \cup C_{3\rightarrow1}$ according to Corollary 3, where

$(Q_{(\omega2, 3)})_{t3} = Q_{(\omega2', 3)}$, $(\omega_2', 3)$: $m(p_1)+m(p_3)+m(p_5) \leq 3$ and

$(Q_{(\omega3, 3)})_{t3} = Q_{(\omega3', 3)}$, $(\omega_3', 3)$: $m(p_1)+m(p_4)+m(p_5) \leq 3$.

For convenience, let

$a=m(p_1)+m(p_2)+m(p_5)$;

$b=m(p_1)+m(p_3)$; and

$c= m(p_1)+m(p_4)$.

Then, we have

$C_{2\rightarrow1}$:

$$\begin{cases} m(p_5) \geq 1 \\ b = 3 \\ a = 4 \end{cases} \vee \begin{cases} m(p_5) \geq 2 \\ b = 2 \\ 3 < a \leq 5 \end{cases} \vee \begin{cases} m(p_5) \geq 3 \\ b = 1 \\ 3 < a \leq 6 \end{cases} \vee \begin{cases} m(p_5) \geq 4 \\ b = 0 \\ 3 < a \leq 7 \end{cases}$$

$C_{3\rightarrow1}$:

$$\begin{cases} m(p_5) \geq 1 \\ c = 3 \\ a = 4 \end{cases} \vee \begin{cases} m(p_5) \geq 2 \\ c = 2 \\ 3 < a \leq 5 \end{cases} \vee \begin{cases} m(p_5) \geq 3 \\ c = 1 \\ 3 < a \leq 6 \end{cases} \vee \begin{cases} m(p_5) \geq 4 \\ c = 0 \\ 3 < a \leq 7 \end{cases}$$

*Remark*: Corollary 3 presents the equivalent transformation of a disjunction of linear constraints via an uncontrollable transition. It directly obtains a transformed result without the analysis on reachability set. Obviously, it is of polynomial complexity.

*Remark*: Corollary 3 shows that the transformed result may be a disjunction of the conjunctions of linear constraints due to the presence of the complementary-marking sets. For such a result, how to perform the follow-up equivalent transformation on it via another transition remains open and thus it is a problem we intend to solve in the future.

### D. Two Rules for Linear Constraint Transformation

In this subsection, two rules to decide the priority of transitions for transformation are presented, which facilitates the transformation from a given linear constraint into a logic expression of linear constraints that can describe the entire admissible-marking set. Note that the transformation procedure is based on Algorithm 1.

Known from Algorithm 1, it is possible that a given linear constraint can be transformed via different transition sequences. Moreover, the transformation via different transition sequences can lead to different computational complexity and different structural complexity of the transformed result. Hence, the priority of transitions for transformation should be considered.

Theorem 7 shows that if a linear constraint is transformed via a transition whose weight is positive, the number of linear constraints in the transformed result is the same as that of the transition's input places. This means the more input places that a transition for transformation has, the more complex the transformed result is. Besides, a complex transformed result can increase the difficulty of the follow-up transformation via anther transition. Hence, Rule 1 is presented for linear constraint transformation.

**Rule 1**: A transition with the fewest input places has the priority for transformation.

Reconsider the net in Fig. 6 with the legal-marking set $\mathcal{L}$: $m(p_1)+m(p_2)+m(p_3) \leq 3$. There are two transition sequences for transformation and they are $\sigma_1 = t_3t_4$ and $\sigma_2 = t_4t_3$, respectively.

According to Rule 1, $\sigma_2$ should be chosen. The transformation procedure via $\sigma_2$ is presented as follows:

$\mathcal{L}_{t4}$: $m(p_1)+m(p_2)+m(p_3)+m(p_5) \leq 3$

$\mathcal{L}_{t4t3}$: $m(p_1)+m(p_2)+m(p_3)+m(p_5)+m(p_4) \leq 3 \vee m(p_1)+m(p_2)+m(p_3)+2m(p_5) \leq 3$

For comparison, we also present the transformation procedure via $\sigma_1$ as follows:

$\mathcal{L}_{t3}$: $m(p_1)+m(p_2)+m(p_3)+2m(p_4) \leq 3 \vee m(p_1)+m(p_2)+m(p_3)+2m(p_5) \leq 3$

$\mathcal{L}_{t3t4}$: $m(p_1)+m(p_2)+m(p_3)+2m(p_4)+m(p_5) \leq 3 \vee m(p_1)+m(p_2)+m(p_3)+2m(p_5) \leq 3 \vee C_{1\rightarrow2}$

Let

$a=m(p_1)+m(p_2)+m(p_3)+2m(p_4)$ and

$b=m(p_1)+m(p_2)+m(p_3)+2m(p_5)$.

Then, we have

$C_{1\rightarrow2}$:

$$\begin{cases} m(p_5) \geq 1 \\ a = 3 \\ b = 4 \end{cases} \vee \begin{cases} m(p_5) \geq 2 \\ a = 2 \\ 3 < b \leq 5 \end{cases} \vee \begin{cases} m(p_5) \geq 3 \\ a = 1 \\ 3 < b \leq 6 \end{cases} \vee \begin{cases} m(p_5) \geq 4 \\ a = 0 \\ 3 < b \leq 7 \end{cases}$$

It is clear that the transformation via $\sigma_2$ is easier than that via $\sigma_1$ and the transformed result via $\sigma_2$ is simpler in structure than that via $\sigma_1$ although they describe the same marking set.

Corollary 3 indicates that for a disjunction of multiple linear constraints, its transformation result via a transition may be a disjunction of the conjunctions of linear constraints due to the presence of the complementary-marking set. For such a result, since the existing methods, as far as we know, cannot perform the follow-up equivalent transformation on it via another transition, a logic expression of linear constraints that describes the entire admissible-marking set may fail to be obtained. Hence, we present Rule 2 to avoid the presence of the complementary-marking sets during the transformation procedure if possible.

**Rule 2**: A transition via which the transformation yields an empty complementary-marking set has the priority.

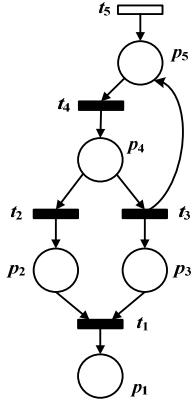

Fig. 11. A Petri net with $\mathcal{L}$: $m(p_1) \leq 1$

Consider the Petri net in Fig. 11 with $T_u = \{t_1 - t_4\}$. $\mathcal{L}$: $m(p_1) \leq 1$ is the legal-marking set for the net. According to Rule 2, $\sigma = t_1 t_2 t_4 t_3 t_4 t_3 t_4$ should be the transition sequence for transformation and the transformation procedure is presented as follows:

$\mathcal{L}_{t1}$: $m(p_1) + m(p_2) \leq 1 \lor m(p_1) + m(p_3) \leq 1$

$\mathcal{L}_{t1t2}$: $m(p_1) + m(p_2) + m(p_4) \leq 1 \lor m(p_1) + m(p_3) \leq 1$

$\mathcal{L}_{t1t2t4}$: $m(p_1) + m(p_2) + m(p_4) + m(p_5) \leq 1 \lor m(p_1) + m(p_3) \leq 1$

$\mathcal{L}_{t1t2t4t3}$: $m(p_1) + m(p_2) + m(p_4) + m(p_5) \leq 1 \lor$
  $m(p_1) + m(p_3) + m(p_4) \leq 1$

$\mathcal{L}_{t1t2t4t3t4}$: $m(p_1) + m(p_2) + m(p_4) + m(p_5) \leq 1 \lor$
  $m(p_1) + m(p_3) + m(p_4) + m(p_5) \leq 1$

$\mathcal{L}_{t1t2t4t3t4t3}$: $m(p_1) + m(p_2) + m(p_4) + m(p_5) \leq 1 \lor$
  $m(p_1) + m(p_3) + 2m(p_4) + m(p_5) \leq 1$

$\mathcal{L}_{t1t2t4t3t4t3t4}$: $m(p_1) + m(p_2) + m(p_4) + m(p_5) \leq 1 \lor$
  $m(p_1) + m(p_3) + 2m(p_4) + 2m(p_5) \leq 1$

Note that the next transformation of $\mathcal{L}_{t1}$ can be performed via $t_2$ or $t_3$. Suppose that we choose $t_3$ for its next transformation. Clearly, $\mathcal{L}_{t1t3}$: $m(p_1) + m(p_2) \leq 1 \lor m(p_1) + m(p_3) + m(p_4) \leq 1$ is obtained, and then it is easy to see that a complementary-marking set can no doubt arise due to the follow-up transformation via $t_2$. Hence, the next transformation of $\mathcal{L}_{t1}$ is performed via $t_2$ instead of $t_3$ according to Rule 2. Similarly, note that the next transformation of $\mathcal{L}_{t1t2}$ can be performed via $t_3$ or $t_4$. Since the next transformation of $\mathcal{L}_{t1t2}$ via

$t_3$ leads to a complementary-marking set while that via $t_4$ cannot, we choose $t_4$ for the next transformation of $\mathcal{L}_{t1t2}$.

*Remark*: Essentially, Rule 1 is also helpful to avoid the presence of the complementary-marking sets. It is clear that Rule 1 can ensure as few linear constraints during transformation as possible, which thereby reduces the possibility to encounter the complementary-marking sets.

*Remark of Section IV*: This section studies the equivalent transformation of a special form of constraints, i.e., a disjunction of linear constraints. The proposed method is essentially based on the theoretical framework of Section III. According to Algorithm 2, the equivalent transformation of a disjunction of linear constraints via an uncontrollable transition is proposed, which corrects the fault in [17]. However, the transformed result may involve the complementary-marking set, which is usually expressed by a complex logic expression of linear constraint whose transformation method is beyond the study scope of our paper. Hence, two rules are thereby proposed to decide the priority of transitions for transformation, which can reduce the possibility that the complementary-marking sets appear during transformation and thus facilitate the transformation from a given linear constraint into a logic expression of linear constraints that can describe the entire admissible-marking set. Unfortunately, for some cases, it is impossible to avoid the presence of complementary-marking sets during a transformation procedure no matter which transition sequence is chosen. Hence, the study on the follow-up equivalent transformation on a transformed result involving complementary-marking sets via other transitions is necessary, which will be studied in our future work.

Although the proposed method has some limitations, it is applicable for many cases. More importantly, we believe that equivalent constraint transformation can be completely solved under the theoretical framework of Section III as long as the equivalent transformation of constraints in any specific form via an uncontrollable transition is proposed.

## V. EXAMPLES

This section presents two examples to illustrate the transformation procedure from a given linear constraint to a logic expression of linear constraints that can describe its entire admissible-marking set.

**Example 1:**

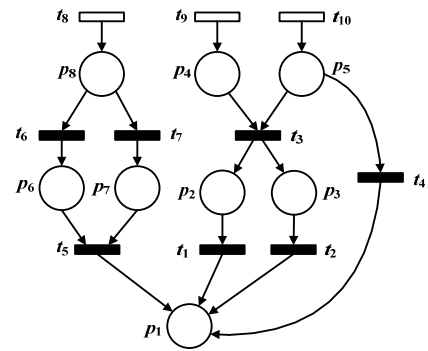

Fig. 12. A Petri net of Example 1

$\mathcal{L}$: $m(p_1) \leq 3$ is the legal-marking set for the Petri net in Fig. 12 whose $T_u = \{t_1 \text{-} t_7\}$.

Based on Algorithm 1, we have $\sigma = t_1 t_2 t_4 t_3 t_5 t_6 t_7$ for transformation according to Rules 1 and 2, resulting in $\mathcal{L}_{t_1 t_2 t_4 t_3 t_5 t_6 t_7} = \mathcal{A}$. The equivalent transformation is performed as follows:

$\mathcal{L}_{t_1}$: $m(p_1) + m(p_2) \leq 3$

$\mathcal{L}_{t_1 t_2}$: $m(p_1) + m(p_2) + m(p_3) \leq 3$

$\mathcal{L}_{t_1 t_2 t_4}$: $m(p_1) + m(p_2) + m(p_3) + m(p_5) \leq 3$

$\mathcal{L}_{t_1 t_2 t_4 t_3}$:
$m(p_1) + m(p_2) + m(p_3) + m(p_5) + m(p_4) \leq 3$ $\lor$
$m(p_1) + m(p_2) + m(p_3) + 2m(p_5) \leq 3$

$\mathcal{L}_{t_1 t_2 t_4 t_3 t_5}$:
$m(p_1) + m(p_2) + m(p_3) + m(p_5) + m(p_4) + m(p_6) \leq 3$ $\lor$
$m(p_1) + m(p_2) + m(p_3) + m(p_5) + m(p_4) + m(p_7) \leq 3$ $\lor$
$m(p_1) + m(p_2) + m(p_3) + 2m(p_5) + m(p_6) \leq 3$ $\lor$
$m(p_1) + m(p_2) + m(p_3) + 2m(p_5) + m(p_7) \leq 3$

$\mathcal{L}_{t_1 t_2 t_4 t_3 t_5 t_6}$:
$m(p_1) + m(p_2) + m(p_3) + m(p_5) + m(p_4) + m(p_6) + m(p_8) \leq 3$ $\lor$
$m(p_1) + m(p_2) + m(p_3) + m(p_5) + m(p_4) + m(p_7) \leq 3$ $\lor$
$m(p_1) + m(p_2) + m(p_3) + 2m(p_5) + m(p_6) + m(p_8) \leq 3$ $\lor$
$m(p_1) + m(p_2) + m(p_3) + 2m(p_5) + m(p_7) \leq 3$

$\mathcal{L}_{t_1 t_2 t_4 t_3 t_5 t_6 t_7}$:
$m(p_1) + m(p_2) + m(p_3) + m(p_5) + m(p_4) + m(p_6) + m(p_8) \leq 3$ $\lor$
$m(p_1) + m(p_2) + m(p_3) + m(p_5) + m(p_4) + m(p_7) + m(p_8) \leq 3$ $\lor$
$m(p_1) + m(p_2) + m(p_3) + 2m(p_5) + m(p_6) + m(p_8) \leq 3$ $\lor$
$m(p_1) + m(p_2) + m(p_3) + 2m(p_5) + m(p_7) + m(p_8) \leq 3$ $\lor$
$C_{2 \to 1}$ $\lor$ $C_{2 \to 3}$ $\lor$ $C_{4 \to 1}$ $\lor$ $C_{4 \to 3}$

For convenience, let
$a = m(p_1) + m(p_2) + m(p_3) + m(p_5) + m(p_4) + m(p_6) + m(p_8)$;
$b = m(p_1) + m(p_2) + m(p_3) + m(p_5) + m(p_4) + m(p_7) + m(p_8)$;
$c = m(p_1) + m(p_2) + m(p_3) + 2m(p_5) + m(p_6) + m(p_8)$; and
$d = m(p_1) + m(p_2) + m(p_3) + 2m(p_5) + m(p_7) + m(p_8)$.

Then, we have
$C_{2 \to 1}$:
$$\begin{cases} m(p_8) \geq 1 \\ a = 4 \\ b = 3 \end{cases} \lor \begin{cases} m(p_8) \geq 2 \\ 3 < a \leq 5 \\ b = 2 \end{cases} \lor \begin{cases} m(p_8) \geq 3 \\ 3 < a \leq 6 \\ b = 1 \end{cases} \lor \begin{cases} m(p_8) \geq 4 \\ 3 < a \leq 7 \\ b = 0 \end{cases}$$

$C_{2 \to 3}$:
$$\begin{cases} m(p_8) \geq 1 \\ c = 4 \\ b = 3 \end{cases} \lor \begin{cases} m(p_8) \geq 2 \\ 3 < c \leq 5 \\ b = 2 \end{cases} \lor \begin{cases} m(p_8) \geq 3 \\ 3 < c \leq 6 \\ b = 1 \end{cases} \lor \begin{cases} m(p_8) \geq 4 \\ 3 < c \leq 7 \\ b = 0 \end{cases}$$

$C_{4 \to 1}$:
$$\begin{cases} m(p_8) \geq 1 \\ a = 4 \\ d = 3 \end{cases} \lor \begin{cases} m(p_8) \geq 2 \\ 3 < a \leq 5 \\ d = 2 \end{cases} \lor \begin{cases} m(p_8) \geq 3 \\ 3 < a \leq 6 \\ d = 1 \end{cases} \lor \begin{cases} m(p_8) \geq 4 \\ 3 < a \leq 7 \\ d = 0 \end{cases}$$

$C_{4 \to 3}$:
$$\begin{cases} m(p_8) \geq 1 \\ c = 4 \\ d = 3 \end{cases} \lor \begin{cases} m(p_8) \geq 2 \\ 3 < c \leq 5 \\ d = 2 \end{cases} \lor \begin{cases} m(p_8) \geq 3 \\ 3 < c \leq 6 \\ d = 1 \end{cases} \lor \begin{cases} m(p_8) \geq 4 \\ 3 < c \leq 7 \\ d = 0 \end{cases}$$

**Example 2:**

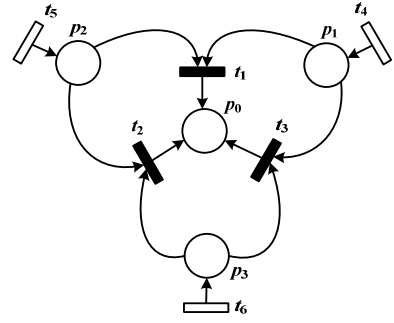

Fig. 13. A Petri net of Example 2

$\mathcal{L}$: $m(p_0) \leq 1$ is the legal-marking set for the Petri net in Fig. 13 whose $T_u = \{t_1 \text{-} t_3\}$.

We choose $\sigma = t_1 t_2 t_3$ for transformation and $\mathcal{L}_{t_1 t_2 t_3} = \mathcal{A}$ holds. The equivalent transformation is performed as follows:

$\mathcal{L}_{t_1}$: $m(p_0) + m(p_1) \leq 1$ $\lor$ $m(p_0) + m(p_2) \leq 1$

$\mathcal{L}_{t_1 t_2}$:
$m(p_0) + m(p_1) + m(p_2) \leq 1$ $\lor$
$m(p_0) + m(p_1) + m(p_3) \leq 1$ $\lor$
$m(p_0) + m(p_2) \leq 1$

$\mathcal{L}_{t_1 t_2 t_3}$:
$m(p_0) + m(p_1) + m(p_2) \leq 1$ $\lor$
$m(p_0) + m(p_1) + m(p_3) \leq 1$ $\lor$
$m(p_0) + m(p_2) + m(p_1) \leq 1$ $\lor$
$m(p_0) + m(p_2) + m(p_3) \leq 1$ $\lor$ $C_{2 \to 3}$

We have
$C_{2 \to 3}$:
$$\begin{cases} m(p_1) \geq 1 \\ m(p_3) \geq 1 \\ m(p_0) + m(p_2) = 1 \\ m(p_0) + m(p_1) + m(p_3) = 2 \end{cases} \lor \begin{cases} m(p_1) \geq 2 \\ m(p_3) \geq 2 \\ m(p_0) + m(p_2) = 0 \\ 1 < m(p_0) + m(p_1) + m(p_3) \leq 3 \end{cases}$$

*Remark:* To our best knowledge, the proposed method is the first one that can perform the equivalent transformation for Examples 1 and 2 without the analysis on reachability set.

It is known that constraint transformation methods based on the analysis on reachability set [2]-[8], [10] can usually obtain an equivalent transformed result. However, they can hardly be applied to large-sized nets due to the state explosion problem. As for the existing equivalent transformation methods without the analysis on reachability set like [18]-[20], [26], they are applicable to a specific class of Petri nets only and fail to perform the equivalent transformation for Examples 1 and 2.

Hence, the proposed method is usually superior to the methods requiring the analysis on reachability set in terms of computational efficiency and memory consumption and it has wider application scope than the existing methods without the analysis on reachability set.

## VI. Conclusion

The major contribution of this work is the new results that reveal the equivalent transformation from a disjunction of multiple linear constraints into a new logic expression of linear constraints via an uncontrollable transition. Based on it, two rules are given for transforming a given linear constraint into a logic expression of linear constraints that can describe its entire admissible-marking set. However, not all the given linear constraints can be successfully transformed based on these two rules. This is because the complementary-marking set may appear during transformation, which is usually expressed by complex logic expression of linear constraints beyond the study scope of our paper. Hence, our future work intends to study transformation methods for more general logic expressions of linear constraints and propose a method that can perform the follow-up equivalent transformation on a transformed result involving complementary-marking sets via other transitions.